\documentclass[11pt,showpacs,preprintnumbers,superscriptaddress,amsmath,amssymb,nofootinbib]{revtex4}
\usepackage{graphicx}% Include figure files
\usepackage{dcolumn}% Align table columns on decimal point
\usepackage{bm}% bold math
\usepackage{amssymb}
\usepackage{amsmath}
\usepackage{epsfig}    
\usepackage{color}
\usepackage{slashed}
\usepackage{hhline}
\usepackage{ulem}
\def\be{\begin{equation}}
\def\ee{\end{equation}}
\newcommand{\bea}{\begin{eqnarray}}
\newcommand{\eea}{\end{eqnarray}}
\newcommand{\nn}{\nonumber}

\begin{document}

%{\begin{flushright}{APCTP Pre2023 - 0XX}\end{flushright}}

%%%%%%%%%
\title{A more novel approach of radiative linear seesaw in a modular $A_4$ symmetry}

%\preprint{KYUSHU-HET-268}

\author{Takaaki Nomura}
\email{nomura@scu.edu.cn}
\affiliation{College of Physics, Sichuan University, Chengdu 610065, China}

\author{Hiroshi Okada}
\email{hiroshi3okada@htu.edu.cn}
\affiliation{Department of Physics, Henan Normal University, Xinxiang 453007, China}

\date{\today}

\begin{abstract}
{
We propose a radiatively induced linear seesaw model that is perfectly realized at leading order by a modular $A_4$ symmetry.
Because this group also has a flavor symmetry, we obtain some predictions concentrating on two specific regions at nearby fixed points $\tau=\omega$ and $i$.  
Through our numerical chi square analysis, we show predictions for each the case depending on the neutrino mass ordering; normal and inverted one.
}
%$A_4$
%%%%%%%%%%%%%%%%%%%%%%%%%%
 %
 \end{abstract}
\maketitle
\newpage

\section{Introduction}
Understanding the neutrino mass structure is one of our important tasks beyond the standard model(BSM).
Although a renown model would be a canonical seesaw mechanism~\cite{Yanagida:1979gs, Minkowski:1977sc,Mohapatra:1979ia,Zee:1980ai} introducing heavy right-handed neutral fermions, it would be difficult to test this model since it demands much higher energy scale than the current experimental energy scale, i.e. $\sim$TeV.
Linear seesaw mechanism~\cite{Wyler:1982dd, Akhmedov:1995ip, Akhmedov:1995vm} is well-known as providing a verifiable model in current experiments at TeV scale %beyond the standard model (BSM), 
introducing left-handed $S_L$ and right-handed $N_R$ neutral fermions~\footnote{An inverse seesaw mechanism also requests these neutral fermions and leads to a similar motivation of linear seesaw~\cite{Wyler:1982dd, Mohapatra:1986bd}.}.
In basis of $[\nu_L,N^C_R,S_L]$, the mass matrix of neutral fermions to realize the linear seesaw is given by
\begin{align}
 \left(\begin{array}{ccc} 
0 & m_D^T &  m'^T_D \\
m_D & 0 & \mu_{NS}  \\
{m'_D} & \mu_{NS}^T & 0 \end{array} \right).
\label{eq:lms}
\end{align}
Assuming the following mass hierarchies
\begin{align}
  m'_D \ll m_D \lesssim \mu_{NS} ,\label{eq:lcond}
\end{align}
we find the following active neutrino mass matrix
\begin{align}
(m'_D)^T (\mu_{NS}^T)^{-1} m_D + (m_D)^T (\mu_{NS})^{-1}  m'_D.
\end{align}
%%%
In a view point of model building, 
most of models put the condition, especially $m'_D \ll m_D$, in Eq.~(\ref{eq:lcond}) by hand.
There exit some ways to realize this condition. Below, we review two examples.
The first example is to introduce $SU(2)_L$ multiplet particles (larger than triplet) that have nonzero vacuum expectation value (VEV) $v'$ 
and generate $m'_D$ via this VEV after the spontaneous symmetry breaking.
Since this VEV receives a constraint $v' { \lesssim} {\cal O}(1)$ GeV from electroweak precision test, $m'_D \ll m_D$ would be realized when $m_D$ is obtained via VEV of the SM Higgs. 
In ref.~\cite{Nomura:2018ibs}, $m'_D$ is obtained via VEV of $SU(2)_L$ quartet Higgs boson.
Also hidden Abelian gauge symmetry is introduced in order to realize the mass matrix in Eq~(\ref{eq:lms}). 
The second example is to generate $m'_D$ at loop level by introducing new particles.
In ref.~\cite{Das:2017ski}, $m'_D$ is induced at one-loop level in which singly-charged heavier leptons and two types of singly-charged bosons run.
In addition, an Abelian gauge $B-L$ symmetry and discrete symmetry $Z_2$ are introduced to realize the model. 

In this paper, we construct a one-loop induced linear seesaw mechanism revisiting our previous paper~\cite{Das:2017ski}.
In fact, this paper has unsatisfactory point; %one short coming.
it includes an inverse seesaw mechanism as a leading order as well as linear seesaw.
Therefore, we could not completely control the desired mass matrix in Eq.~(\ref{eq:lms}) via introduced two additional symmetries.
In order to resolve this defect, we introduce a modular $A_4$ symmetry.
Surprisingly, this symmetry solely realizes {linear seesaw mechanism} and gives us predictions at nearby the fixed points $\tau= \{ \omega,\ i \}$ in both cases of neutrino mass hierarchies.~\footnote{There exists wide variety of applications to explain lepton and quark masses and mixings as well as new physics~\cite{Feruglio:2017spp, Criado:2018thu, Kobayashi:2018scp, Okada:2018yrn, Nomura:2019jxj, Okada:2019uoy, deAnda:2018ecu, Novichkov:2018yse, Nomura:2019yft, Okada:2019mjf,Ding:2019zxk, Nomura:2019lnr,Kobayashi:2019xvz,Asaka:2019vev,Zhang:2019ngf, Gui-JunDing:2019wap,Kobayashi:2019gtp,Nomura:2019xsb, Wang:2019xbo,Okada:2020dmb,Okada:2020rjb, Behera:2020lpd, Behera:2020sfe, Nomura:2020opk, Nomura:2020cog, Asaka:2020tmo, Okada:2020ukr, Nagao:2020snm, Okada:2020brs,Kang:2022psa, Ding:2024fsf, Ding:2023htn, Nomura:2023usj, Kobayashi:2023qzt, Petcov:2024vph, Kobayashi:2023zzc, Nomura:2024ghc, Qu:2024rns, Nomura:2024ctl, Nomura:2024vzw}. } 
Through our numerical chi square analysis, we demonstrate several predictions later.

%%%%%%
This paper is organized as follows.
In Sec. \ref{sec:II}, 
we review our setup of the modular $A_4$ assignments introducing new particles and explain how to realize our scenario by showing the charged-fermion sector and the neutral fermion sector. 
In Sec. \ref{sec:III}, we demonstrate chi-square numerical analyses concentrating on two fixed points and show what kinds of predictions we obtain.
Finally, we summarize and conclude in Sec. \ref{sec:IV}.

\section{Model setup}
\label{sec:II}

\begin{table}[t!]
\begin{tabular}{|c||c|c|c|c|c|c||c|c|c|}\hline\hline  
& ~$L_L$~ & ~$\overline{\ell_R}$~ & ~$\overline{N_R}$~ & ~${S_L}$~ & ~${E_L}$~ & ~$\overline{E_R}$~
& ~$H$~ & ~$\eta$~ & ~$\chi^+$~    \\\hline\hline 
%%%
$SU(2)_L$   & $\bm{2}$  & $\bm{1}$  & $\bm{1}$ & $\bm{1}$ & $\bm{1}$  & $\bm{1}$    & $\bm{2}$& $\bm{2}$ & $\bm{1}$   \\\hline 
$U(1)_Y$    & $-\frac12$  & $1$  & $0$ & $0$ & $-1$  & $+1$ & $\frac12$ & $-\frac12$ & $+1$     \\\hline
$A_4$   & $\bm{3}$  & $ \{ \bar{\bm{1}} \}$ & $\{ \bar{\bm{1}} \}$ & $\{ \bm{1} \} $ & $\{ {\bm{1}} \}$ & $\{ \bar{\bm{1}} \} $  & $\bm{1}$ & $\bm{1}$ & $\bm{1}$          \\\hline 
$-k_I$    & $-2$  & $0$ & $-2$ & $+2$ & $-1$ & $-3$& $0$  & $+1$ & $-1$   \\\hline
\end{tabular}
\caption{Charge assignments of the fermions and bosons
under $SU(2)_L\otimes U(1)_Y \otimes A_4$ where $-k_I$ is the number of modular weight. Here, {$\{ \bm{1} \} =\{1, 1', 1''\}$} indicates assignment of $A_4$ singlets.}\label{tab:1}
\end{table}
{In this section we show the model. The particle contents are almost the same as the model in ref.~\cite{Das:2017ski} where we remove one scalar singlet. 
Then new filed contents are neutral singlet fermions $\overline{N_R}$ and $S_L$, charged singlet fermions $E_L$ and $\overline{E_R}$, inert doublet scalar $\eta$ and inert charged scalar $\chi^+$. 
For symmetry of the model, we removed $Z_2 \times U(1)_{B-L}$ from the original model and added modular $A_4$ one.
The field contents and charge assignments of them are summarized in Table~\ref{tab:1}.}
%%%
Under these symmetries, the renormalizable Lagrangian is given by  
\begin{align}
-{\cal L}_\ell &=
 [ \overline{\ell_R}  Y^{(2)}_{\rm 3} L_L {\tilde H}] 
+
[Y^{(4)}_3 \overline{E_R} L_L \eta ]
+
y_{ES} \overline{E^C_L} S_L \chi^+
+ [\mu_E Y^{(4)}_{1,1'} \overline{E_R} E_L] \nn\\
%
%+{\color{red}y'_{ES} \overline{E_R} S_L \chi^- }
&+
[Y^{(4)}_3 \overline{N_R} L_L H ] 
+
[\mu_N Y^{(4)}_{1,1'} \overline{N_R} N_R^C] 
+
\mu_{NS}  \overline{N_R}S_L \nn\\
&
+ [g' Y^{(4)}_{1,1'}  \overline{\ell_R} E_L \chi_0]
+
{\rm h.c.},
\label{eq:lpy}
\end{align}
where $[\cdots]$ is a trivial $A_4$ singlet ${\bm 1}$ abbreviating free parameters that are explicitly demonstrated later.
$Y^{(2)}_3\equiv [y_1,y_2,y_3]^T$, $Y^{(4)}_3\equiv[y'_1,y'_2,y'_3]^T= [y_1^2-y_2y_3,y_3^2-y_1y_2,y_2^2-y_1y_3]^T$, $Y^{(4)}_1\equiv y^2_1-2y_2y_3$, and $Y^{(4)}_{1'}\equiv y^2_3-2y_1y_2$ are modular couplings that are uniquely fixed when modulus $\tau$ is determined~\cite{Feruglio:2017spp}. In our paper, we concentrate on specific region at nearby fixed points $\tau=i,\omega$.
\\
Renormalizable Higgs potential is found as
\begin{align}
{\cal V} &=
-\mu_H^2 |H|^2 - \mu_\Delta^2 |\eta|^2 - \mu_{\chi^+}^2|\chi^+|^2 - \mu_{\chi_0}^2|\chi_0|^2
+\mu (H^\dag (i\tau_2) \eta \chi^+ +{\rm h.c.})+\mu' (H^T \eta \chi_0 +{\rm h.c.})\nn\\
&+{\rm quartic\ terms},
\label{eq:pot}
\end{align}
where $\tau_2$ is  a second Pauli matrix and several free parameters include modular forms in mass and non-dimensional parameters.

\if0
Here, we mention a little important issue on our model.
We work on non-SUSY scenario that seems to be non-holomorphic theory.
In this case, Feruglio-type modular form would not work but Gui-Jun model might work well.
Theoretically, however, no one knows whether our model would work in the  Feruglio-type.
%%%
Even though the Feruglio-type would not work in non-SUSY model, our scenario can transfer to the SUSY scenario without violating any predictions our models.
(Note that there is a problem only when a model includes non trivial quartic term in the potential such as Ma model $(H^\dag \eta)^2$.)
This is why we simply work on non-SUSY model. 
\fi

\subsection{Charged-lepton mass matrix}
The mass matrix of charged-lepton comes from the first term in Eq.~(\ref{eq:lpy}) after spontaneous electroweak symmetry breaking and it is found as
\begin{align}
m_\ell 
= \frac{v}{\sqrt2}
 \left(\begin{array}{ccc} 
a_e & 0 & 0 \\
0 & b_e& 0\\
0& 0 & c_e  \end{array} \right)
 \left(\begin{array}{ccc} 
y_1 & y_3 & y_2 \\
y_2 & y_1 & y_3\\
y_3& y_2 & y_1  \end{array} \right),
\label{eq:cgd-mtx}
\end{align}
where $\{a_e, b_e, c_e \}$ can be real parameters without loss of generality.
$m_\ell$ is diagonalized by $D_\ell=V_{e_R}^\dag m_\ell V_{e_L}$. 
Therefore, $|D_\ell|^2=V^\dag_{e_L} m^\dag_\ell m_\ell V_{e_L}$ where $D_\ell=$
diag.$(m_e,m_\mu,m_\tau)$.
%%%
\if0
{\color{red} %この説明はなくても良いかもしれない。
Since free parameters to fix the charged-lepton masses $\{a_e, b_e, c_e \}$ do not mix with modular Yukawa parameters $\{y_1, y_2, y_3 \}$,
we can directly solve these free parameters as follows.}
\fi
%%%
From $|D_\ell|^2=V^\dag_{e_L} m^\dag_\ell m_\ell V_{e_L}$, $\{a_e, b_e, c_e \}$ are fixed to fit the three observed charged-lepton masses.
% by the following relations:
\begin{align}
&{\rm Tr}[m^\dag_\ell m_\ell] = |m_e|^2 + |m_\mu|^2 + |m_\tau|^2 = \frac{v^2}2 (a_e^2+b_e^2+c_e^2)( |y_1|^2+|y_2|^2+|y_3|^2),\\
& {\rm Det}[m^\dag_\ell m_\ell] = |m_e|^2  |m_\mu|^2  |m_\tau|^2
= a_e^2b_e^2c_e^2 |y_1^3+y_2^3+y_3^3-3y_1y_2y_3|^2, \\
&\frac{({\rm Tr}[m^\dag_\ell m_\ell])^2 -{\rm Tr}[(m^\dag_\ell m_\ell)^2]}2 = |m_e|^2  |m_\mu|^2 + |m_\mu|^2  |m_\tau|^2+ |m_e|^2  |m_\tau|^2
\nn\\
&=\frac{v^6}8 
(a_e^2b_e^2+b_e^2c_e^2+c_e^2a_e^2)[( |y_1|^2+|y_2|^2+|y_3|^2)^2-|y_1 y_3^*+y_2 y_1^*+y_3 y_2^*|^2],
\label{eq:l-cond0}
\end{align}
therefore combinations $a_e,b_e,c_e$ can be rewritten in terms of three charged-lepton mass eigenvalues and $y_{1,2,3}$ as follows:
\begin{align}
& A\equiv a_e^2+b_e^2+c_e^2 = \frac{v^2}2 \frac{ |y_1|^2+|y_2|^2+|y_3|^2}{ |m_e|^2 + |m_\mu|^2 + |m_\tau|^2},\\
&C\equiv a_e^2b_e^2c_e^2
= \frac{ |y_1^3+y_2^3+y_3^3-3y_1y_2y_3|^2}{|m_e|^2  |m_\mu|^2  |m_\tau|^2}, \\
& B\equiv a_e^2b_e^2+b_e^2c_e^2+c_e^2a_e^2
=\frac{v^6}8 
\frac{( |y_1|^2+|y_2|^2+|y_3|^2)^2-|y_1 y_3^*+y_2 y_1^*+y_3 y_2^*|^2}{|m_e|^2  |m_\mu|^2 + |m_\mu|^2  |m_\tau|^2+ |m_e|^2  |m_\tau|^2}.
\label{eq:l-cond0}
\end{align}
It suggests that $a_e^2, b_e^2,c_e^2$ are obtained as a result of solution of the following equation:
\begin{align}
x^3 -A x^2 + B x - C =0.
\end{align}
Then, the tadpole condition;
%the stationary conditions; 
$3x^2-2Ax+B=0$, gives the following solutions 
\begin{align}
x_\pm = \frac{A\pm \sqrt{A^2-3B}}{3}. \label{eq:statcnd}
\end{align}
Since all solutions $x_1< x_2 < x_3$ have to be positive reals and Eq.~(\ref{eq:statcnd}) simply increases, we need the following conditions:
\begin{align}
0< x_1 < x_-,\quad x_- <  x_2 < x_+,\quad x_+ < x_3,
\end{align}
where $x_i$ (i$=$1,2,3) corresponds to $a_e^2,b_e^2,c_e^2$, depending on hierarchies among these parameters.
% 数値計算で初めてx1がae^2, be^2,ce^2のどれになるかが決まる。ちょっとややこしい書き方だけど正確に書こうとするとこうなる。
These conditions are helpful to have our numerical analysis.

\subsection{Heavier charged-lepton mass matrix}
The mass matrix of heavier charged-lepton comes from the $\mu_E$ term in Eq.~(\ref{eq:lpy}) as follows:
\begin{align}
M_E 
= \frac{v}{\sqrt2}
 \left(\begin{array}{ccc} 
|m_1| & 0 & m_2 \\
m_4 & |m_3| & 0\\
0 & m_6 & |m_5|  \end{array} \right),
\label{eq:cgd-mtx}
\end{align}
where $\{|m_1|,|m_3|,|m_5| \}$ can be real parameters without loss of generality while the others are complex values.
$M_E$ is diagonalized by $D_E=V_{E_R}^\dag M_E V_{E_L}$. 
Therefore, $|D_E|^2=V^\dag_{E_L} M_E^\dag M_E V_{E_L}$.
Flavor eigenstate $(E_{L,R})$ are written in terms of mass eigenstate $(\psi_{L,R})$ and their mixing matrix $(V_{E_{L,R}})$ as follows: $E_{L,R}= V_{E_{L,R}} \psi_{L,R}$.

\subsection{Neutral fermion mass matrix \label{sec:nuetral}}
%In this subsection, 
The mass matrix of neutral fermions comes from the following terms
after spontaneous electroweak symmetry breaking;
\begin{align}
 &
[Y^{(4)}_3 \overline{N_R} L_L H ] 
+
[\mu_N Y^{(4)}_{1,1'} \overline{N_R} N_R^C] 
+
\mu_S  \overline{N_R}S_L \nn\\
&\hspace{1cm}+[Y^{(4)}_3 \overline{E_R} L_L \eta ]
+
y_{ES} \overline{E^C_L} S_L \chi^+
+ 
[\mu_E Y^{(4)}_{1,1'} \overline{E_R} E_L]
+
{\rm h.c.}\nn\\
&\supset
m_D \overline{N_R} \nu_L 
+
M_R \overline{N_R} N_R^C
+
\mu_{NS}  \overline{N_R}S_L \nn\\
&
\hspace{1cm}+\overline{E_R} f' \nu_L \eta^+
+
y_{ES} \overline{E^C_L} S_L \chi^+
+ 
M_E \overline{E_R} E_L
-\mu H^\dag (i\tau_2) \eta \chi^+
+
{\rm h.c.}
\label{eq:neut}
\end{align}
In basis of $[\nu_L, {N^C_R},S_L]^T$, the neutral fermion mass matrix is given by
\begin{align}
{\cal M_N}
&
=
 \left(\begin{array}{ccc} 
0 & m_D^T & \delta m'^T_D \\
m_D & M_N & \mu_{NS}  \\
\delta m'_D & \mu_{NS}^T & 0 \end{array} \right) ,
\label{eq:nfm}
\end{align}
where each of mass matrix is given by
\begin{align}
m_D
&
=
\frac{v \alpha_N}{\sqrt2}
 \left(\begin{array}{ccc} 
1 & 0 & 0 \\
0 & \beta_N & 0  \\
0 &0 & \gamma_N \end{array} \right) 
 \left(\begin{array}{ccc} 
y'_1 & y'_3 & y'_2 \\
y'_2 & y'_1 & y'_3 \\
y'_3 & y'_2 & y'_1  \end{array} \right) \equiv \frac{v \alpha_N}{\sqrt2} \widetilde m_D,\nn\\
% %%
M_N
&=
\mu_N
 \left(\begin{array}{ccc} 
1 & \widetilde\mu_{N_{12}} & 0 \\
\widetilde \mu_{N_{12}} & 0 &  \widetilde\mu_{N_{23}}  \\
0 &  \widetilde\mu_{N_{23}} & \widetilde \mu_{N_{33}} \end{array} \right) 
 \equiv \mu_N  \widetilde M_N ,\nn\\
% ,
\mu_{NS}
&=
\mu_{NS_1}
 \left(\begin{array}{ccc} 
1 & 0 & 0 \\
 0 & \widetilde \mu_{NS_2} &  0  \\
0 &  0 &\widetilde \mu_{NS_3} \end{array} \right) \equiv \mu_{NS_1}  \widetilde \mu_{NS} .
\label{eq:nfm}
\end{align}
%%%
$\delta m'_D$ is given at one-loop level through the second line in Eq.~(\ref{eq:lpy}) as follows:
\begin{align}
\delta m'_D &=-\frac{\mu v y_{ES_1} \alpha_E}{\sqrt{2}(4\pi)^2} f''^T_{ia} D_{E_a} {(\widetilde Y_{ES})}_{aj}
F_I(D_{E_a} , m_{\chi^\pm},m_{\eta^\pm})
\equiv \mu \delta \widetilde m'_D,\\
F_I(D_{E_a} , m_{\chi^\pm},m_{\eta^\pm}) &=
\int dx dy dz\delta(1-x-y-z) \frac{1}{x D_{E_a}^2 + m_{\chi^\pm}^2+z m_{\eta^\pm}^2},\\
f''&= V^\dag_{E_R} 
 \left(\begin{array}{ccc} 
1 & 0 & 0 \\
0 & \widetilde\beta_E & 0  \\
0 &0 &\widetilde \gamma_E \end{array} \right) 
 \left(\begin{array}{ccc} 
y'_1 & y'_3 & y'_2 \\
y'_2 & y'_1 & y'_3 \\
y'_3 & y'_2 & y'_1  \end{array} \right)  \equiv V^\dag_{E_R} f',\\
\widetilde Y_{ES}&= V^T_{E_L} 
 \left(\begin{array}{ccc} 
1 & 0 & 0 \\
0 & \widetilde y_{ES_2} & 0  \\
0 &0 & \widetilde y_{ES_3} \end{array} \right) \equiv V^T_{E_L} y_{ES}.
\label{eq:mm}
\end{align}
Since $\delta m'_D$ is induced at one-loop level, we can naturally suppose $\delta m'_D$ to be the minimum mass scale among these mass matrices. $m_D$ is proportional to $v$ that is the electroweak scale $\sim100$ GeV, while $M_N$ and $\mu_{NS}$ are directly generated by the bare mass Lagrangian.
Therefore we expect  $M_N$ and $\mu_{NS}$ scale to be $\sim$1 TeV that originates from our expected new physics beyond the SM.
%%%
Moreover, we have a non-unitary condition $\mu_{NS}^{-1} m_D^T\lesssim 10^{-3}$ that comes from several {experimental constraints related to electroweak observables such as}
the SM W boson mass , the effective Weinberg angle, several ratios of Z boson fermionic decays, invisible decay of Z, electroweak universality, measured CKM, and lepton flavor violations. Therefore  $m_D \ll \mu_{NS}$ is required both experimentally and theoretically.
%%%
Hence, the following mass ordering is naturally realized:
\begin{align}
\delta m'_D \ll m_D < M_N,\ \mu_{NS}.
\label{eq:massorder}
\end{align}
Under these mass hierarchies, the active neutrino mass matrix at leading order is given by~\footnote{Although we have the inverse seesaw mass matrix $(\delta m'_D)^T (\mu_{NS})^{-1}  M_N (\mu_{NS})^{-1} \delta m'_D$, this term is next leading order due to the hierarchies in Eq.~(\ref{eq:massorder}). }
\begin{align}
m_\nu = &- (\delta m'_D)^T (\mu_{NS})^{-1} m_D -(m_D)^T (\mu_{NS})^{-1} \delta m'_D\nn\\
&=
- \mu\frac{v \alpha_N}{\sqrt2 \mu_{NS_1}}
\left[(\delta\widetilde m'_D)^T (\widetilde\mu_{NS})^{-1} \widetilde m_D + (\widetilde m_D)^T (\widetilde \mu_{NS})^{-1} \delta \widetilde m'_D\right]\nn\\
&\equiv \epsilon_\nu \widetilde m_\nu.
% & \hspace{1cm}- (\delta m'_D)^T (\mu_{NS})^{-1}  M_N (\mu_{NS})^{-1} \delta m'_D.
% 
\label{eq:neutmass}
\end{align}
The overall mass parameter $\epsilon_\nu\equiv- \mu\frac{v \alpha_N}{\sqrt2 \mu_{NS_1}}$ is determined by the
atmospheric neutrino mass-squared difference $\Delta m_{\rm atm}^2$ and dimensionless neutrino mass eigenvalues:
\begin{align}
({\rm NH}):\  \epsilon_\nu^2= \frac{|\Delta m_{\rm atm}^2|}{\widetilde D_{\nu_3}^2-\widetilde D_{\nu_1}^2},
\quad
({\rm IH}):\  \epsilon_\nu^2= \frac{|\Delta m_{\rm atm}^2|}{\widetilde D_{\nu_2}^2-\widetilde D_{\nu_3}^2},
 \end{align}
where 
NH and IH respectively stand for the normal
and inverted hierarchies. 
The dimensionless mass eigenvalues for the active neutrinos $\widetilde D_\nu \equiv (m_\nu/\epsilon_\nu)= \{\widetilde D_{\nu_1}, \widetilde D_{\nu_2}, \widetilde D_{\nu_3} \}$ are obtained by diagonalizing the mass matrix as $\widetilde D_\nu \equiv V_\nu^T \widetilde m_\nu V_\nu$, therefore $V_\nu^\dag m_\nu^\dag m_\nu V_\nu ={\rm diag.}(|D_{\nu_1}|^2,|D_{\nu_2}|^2,|D_{\nu_3}|^2)$. 
%where $D_{\nu_{1,2,3}}$ are the three neutrino mass eigenvalues.
Then the observed mixing matrix $U$ is defined by $U=V_L^\dag V_\nu$.
Subsequently, the solar neutrino mass-squared difference is fixed to be
\begin{align}
\Delta m_{\rm sol}^2= {\epsilon_\nu^2}({\widetilde D_{\nu_2}^2-\widetilde D_{\nu_1}^2}).
 \end{align}
$\Delta m_{\rm sol}^2$ has to be within the allowed range of the experimental value, in NuFit 5.2~\cite{Esteban:2020cvm}. 
The sum of neutrino mass eigenvalues is given by
\begin{align}
\sum D_\nu=\epsilon_\nu(\widetilde D_{\nu_1}  +\widetilde D_{\nu_2}+\widetilde D_{\nu_3}).
\end{align}
$\sum D_\nu$ is constrained by the minimal standard cosmological model $\sum D_{\nu}\le$ 120 meV by 
$\Lambda$CDM $+\sum D_{\nu_i}$~\cite{Vagnozzi:2017ovm, Planck:2018vyg}.
However, it is weaker if the data is analyzed in the context of extended cosmological models~\cite{ParticleDataGroup:2014cgo}.
%%%
Moreover, DESI and CMB data combination recently provides more stringent upper bound on the sum as $\sum D_{\nu}\le$ 72 meV~\cite{DESI:2024mwx}. 
The effective mass for neutrinoless double beta decay is given by 
\begin{align}
\langle m_{ee}\rangle=\epsilon_\nu |\widetilde D_{\nu_1} \cos^2\theta_{12} \cos^2\theta_{13}+\widetilde D_{\nu_2} \sin^2\theta_{12} \cos^2\theta_{13}e^{i\alpha_{21}}+\widetilde D_{\nu_3} \sin^2\theta_{13}e^{i(\alpha_{31}-2\delta_{CP})}|,
\end{align}
where $\sin\theta_{12,13},\cos\theta_{12,13}$ are mixing of $U$, $\alpha_{2,3}$ are Majorana phases, and $\delta_{CP}$ is Dirac CP phase. 
 $\langle m_{ee} \rangle$ is constrained by the current KamLAND-Zen data measured in future~\cite{KamLAND-Zen:2024eml},
 and the current  upper bound is given by $\langle m_{ee}\rangle<(36-156)$ meV at 90 \% confidence level. 
Here the range of the bound is due to the use of different method in estimating nuclear matrix elements.

\section{Numerical analysis and phenomenology}
\label{sec:III}
In this section we show chi square analysis to fit the neutrino oscillation data where we refer to NuFit 5.2~\cite{Esteban:2020cvm} as we mentioned in the last section,
and Dirac CP phase as well as Majorana phase are considered as an output predictions since any values are allowed by experiments at 3$\sigma$ confidence level (C.L.).
Therefore, we make the use of five reliable observables [$\Delta m_{\rm atm}^2,\ \Delta m_{\rm sol}^2,\ \sin\theta_{12},\ \sin\theta_{23},\ \sin\theta_{13}$].
Moreover, we focus on the region at nearby two fixed points $\tau=\omega$ and $i$.
We randomly select our input parameters in the ranges of $[10^{-5}-0.1]$ for dimensionless values and $[0.1\ {\rm GeV}-10^5\ {\rm GeV}]$ for mass dimension ones respectively
where we take the mass input parameters so that the non-unitarity condition is satisfied as discussed in the subsection \ref{sec:nuetral}.

\subsection{NH}

Here we summarize our findings for NH case.

 %%%%%%%%%%%%%%%%%%%
\begin{figure}[tb]
\begin{center}
\includegraphics[width=50.0mm]{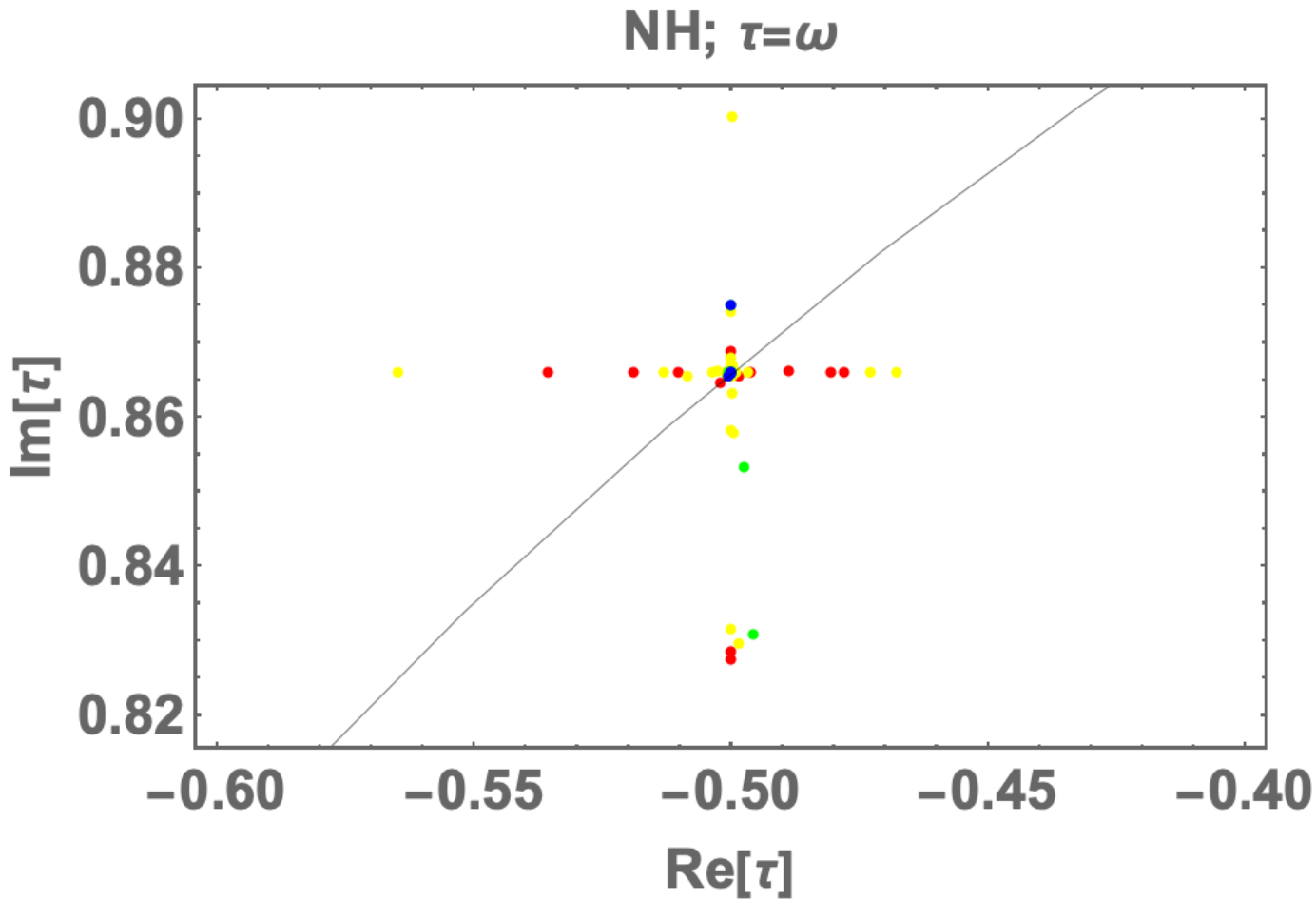} \quad
%%%
%%%
\caption{Allowed region of $\tau$ applying $\Delta\chi^2$ analyses in case of NH at nearby $\tau=\omega$.
{The blue, green, yellow and red points represent the parameter sets that respectively satisfy the range of $0-1$ $\sigma$, $1-2$ $\sigma$, $2-3$ $\sigma$, and $3-5$ $\sigma$ C.L. estimated by chi square analysis,}
where the black solid line is boundary of the fundamental domain at $|\tau|=1$.}
  \label{fig:omega_nh1}
\end{center}\end{figure}
%%%%%%%%%%%%%%%%%%%   
\subsubsection{$\tau=\omega$}
Here, we start to discuss the  case of $\tau=\omega$ in NH.
%In Figs.~\ref{fig:omega_nh1}, we figure out the allowed range of $\tau$ (left), Majorana phases (center), and Dirac CP phase $\delta_{\rm  CP}$ in terms of sum of neutrino masses $\sum D_\nu$.
%
In Fig.~\ref{fig:omega_nh1}, we show allowed region of Re[$\tau$] and Im[$\tau$] at nearby $\tau=\omega$.
{The blue, green, yellow and red points represent the parameter sets that respectively satisfy the range of $0-1$ $\sigma$, $1-2$ $\sigma$, $2-3$ $\sigma$, and $3-5$ $\sigma$ C.L. estimated by chi square analysis,}
where the black solid line is boundary of the fundamental domain at $|\tau|=1$. 
One finds there is deviation only when ${\rm Re}[\omega]\approx0$ or ${\rm Im}[\omega]\approx0$.

 %%%%%%%%%%%%%%%%%%%
\begin{figure}[tb]
\begin{center}
\includegraphics[width=50.0mm]{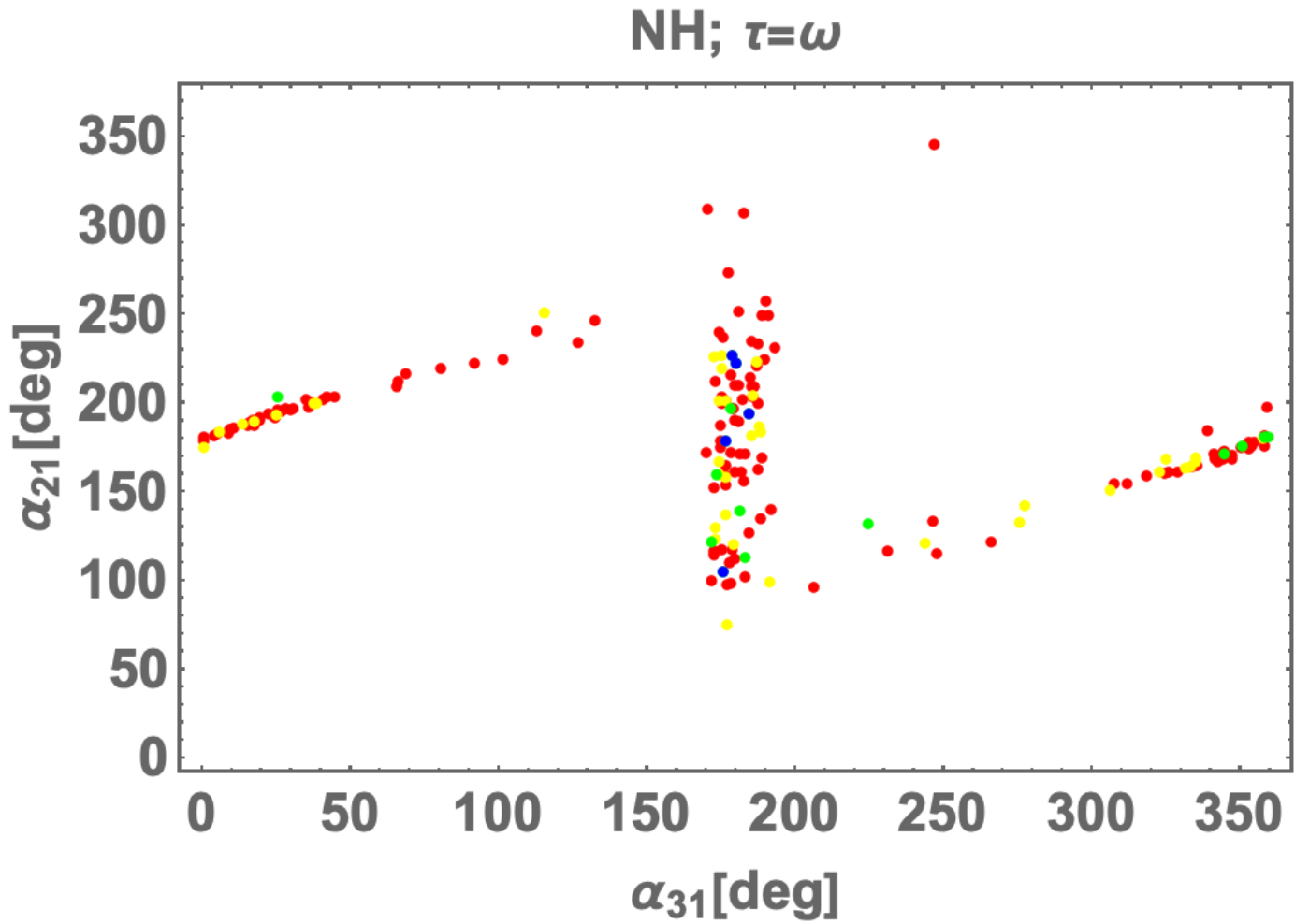} \quad
%%%
\includegraphics[width=50.0mm]{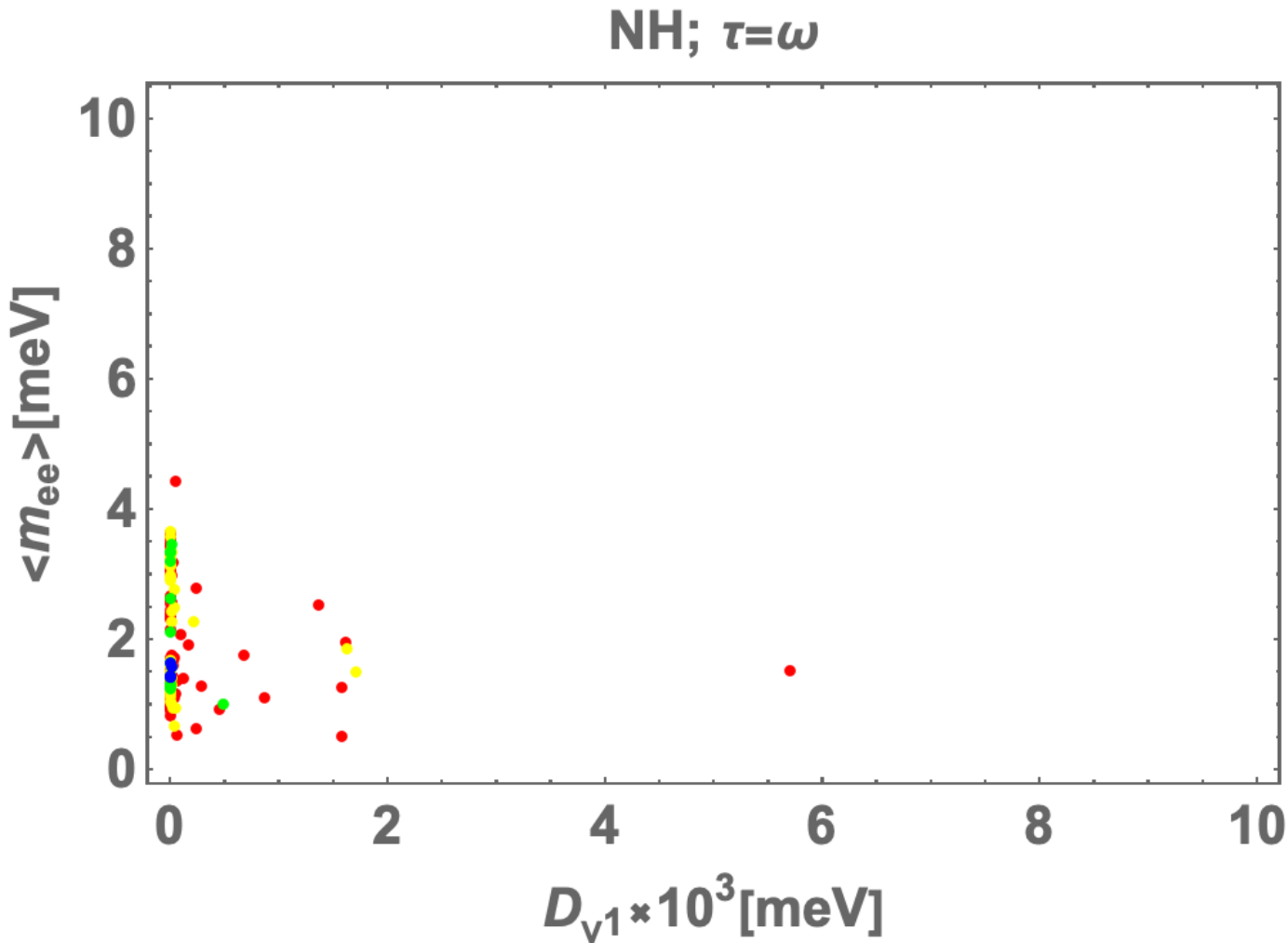} \quad
%%%
\includegraphics[width=50.0mm]{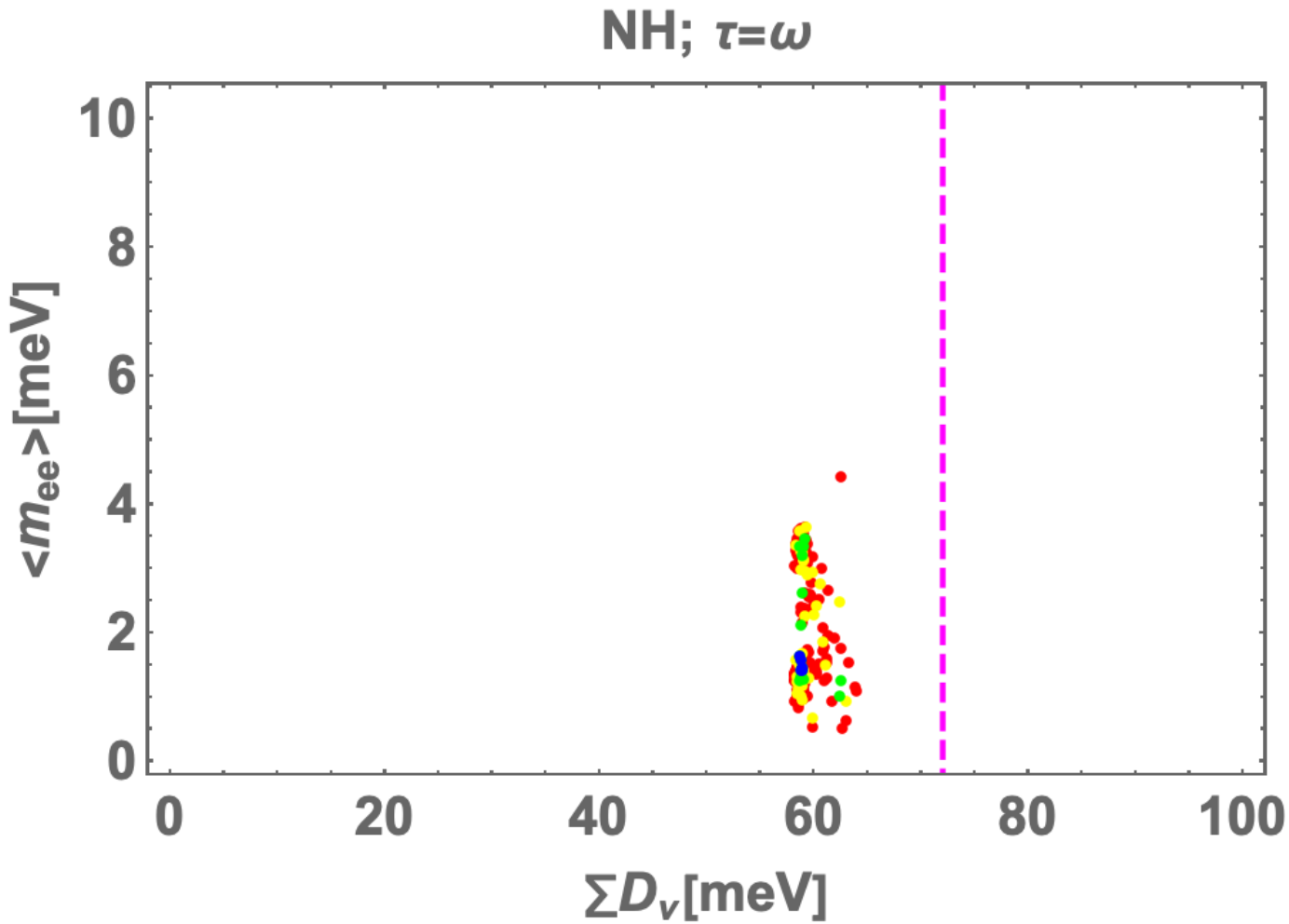} \quad
%%%
\caption{Allowed region of Majorana phases (left), $\langle m_{ee}\rangle$ in terms of the lightest neutrino mass eigenvalue $D_{\nu_1}$ (center), and  $\langle m_{ee}\rangle$ in terms of $\sum D_{\nu}$ (right).
All the legends are the same as the one of Fig.~\ref{fig:omega_nh1}.}
  \label{fig:omega_nh2}
\end{center}\end{figure}
%%%%%%%%%%%%%%%%%%%   
%
In Fig.~\ref{fig:omega_nh2}, we show some allowed regions of Majorana phases (left), $\langle m_{ee}\rangle$ in terms of the lightest neutrino mass eigenvalue $D_{\nu_1}$ (center), and  $\langle m_{ee}\rangle$ in terms of $\sum D_{\nu}$ (right).
All the color legends are the same as the one in Fig.~\ref{fig:omega_nh1}.
The dotted pink vertical line in the right figure shows the upper limit from the combination of the DESI and CMB data that is $\sum m_\nu < 72$ meV.
In the left figure,
we find  $\alpha_{21}$ is  allowed by the ranges [$100^\circ - 300^\circ$].
Although  $\alpha_{31}$ is allowed by whole the ranges, there is a unique correlation between these two Majorana phases.
For example, $\alpha_{31}$ tends to be localized at nearby $180^\circ$ and except this region Majorana phases are uniquely fixed if either of them is determined.
In the center figure,
we found $\langle m_{ee}\rangle=[0.5-4.5]$ [meV] and  $D_{\nu_1}\lesssim6\times10^{-3}$ [meV] both of which are too tiny to test by experiments
where the most of the points are localized at nearby $D_{\nu_1}=0$ meV.
In the right figure,
we found $\sum D_\nu=[58-65]$ [meV] that could be testable in a cosmological experiments/observations such as DESI and CMB data combination in near future. 

\subsubsection{$\tau=i$}

 %%%%%%%%%%%%%%%%%%%
\begin{figure}[tb]
\begin{center}
\includegraphics[width=50.0mm]{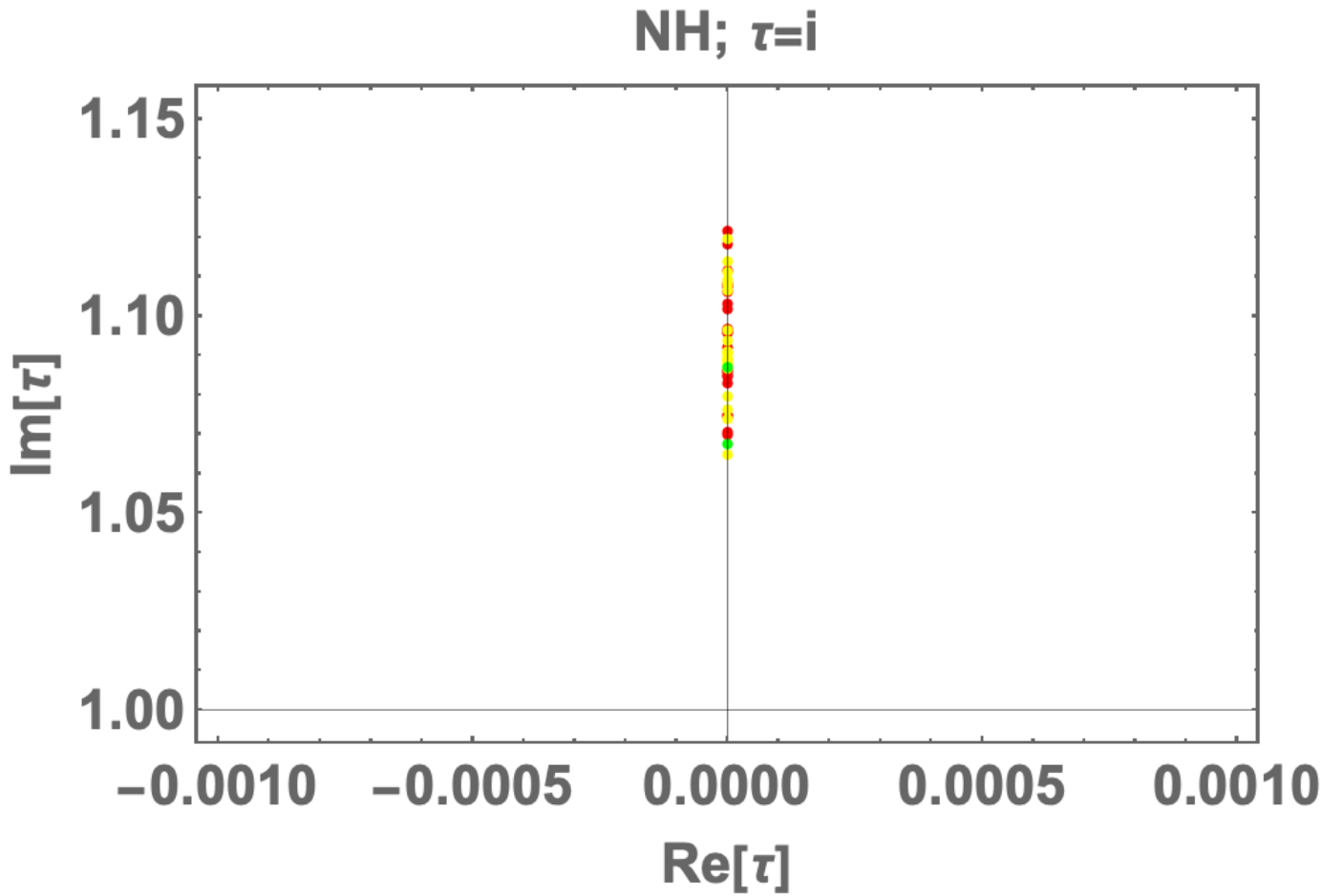} 
%%%
\caption{Allowed region of $\tau$ applying $\Delta\chi^2$ analyses in case of NH at nearby $\tau=\omega$.
All the legends are the same as the one of Fig.~\ref{fig:omega_nh1}.  }
  \label{fig:i_nh1}
\end{center}\end{figure}
%%%%%%%%%%%%%%%%%%%   
%
In Fig.~\ref{fig:i_nh1}, we show allowed region of Re[$\tau$] and Im[$\tau$] at nearby $\tau=i$.
All the color legends are the same as the one of Fig.~\ref{fig:omega_nh1}.
Allowed  range is $\tau=[1.06i\sim 1.12 i]$.
%One finds there is deviation only when ${\rm Re}[\omega]\approx0$ or ${\rm Im}[\omega]\approx0$.

 %%%%%%%%%%%%%%%%%%%
\begin{figure}[tb]
\begin{center}
%%%
\includegraphics[width=50.0mm]{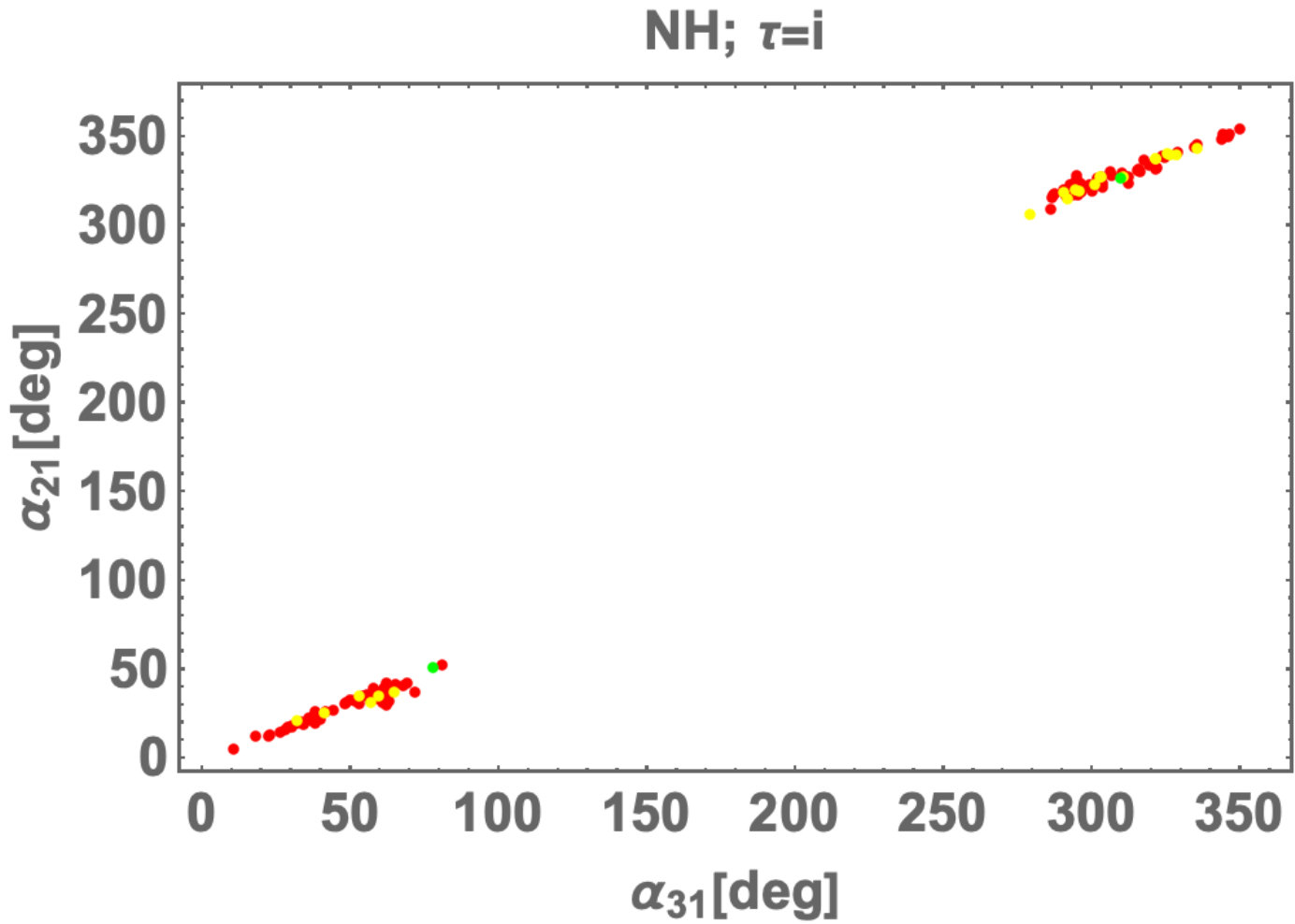} \quad
\includegraphics[width=50.0mm]{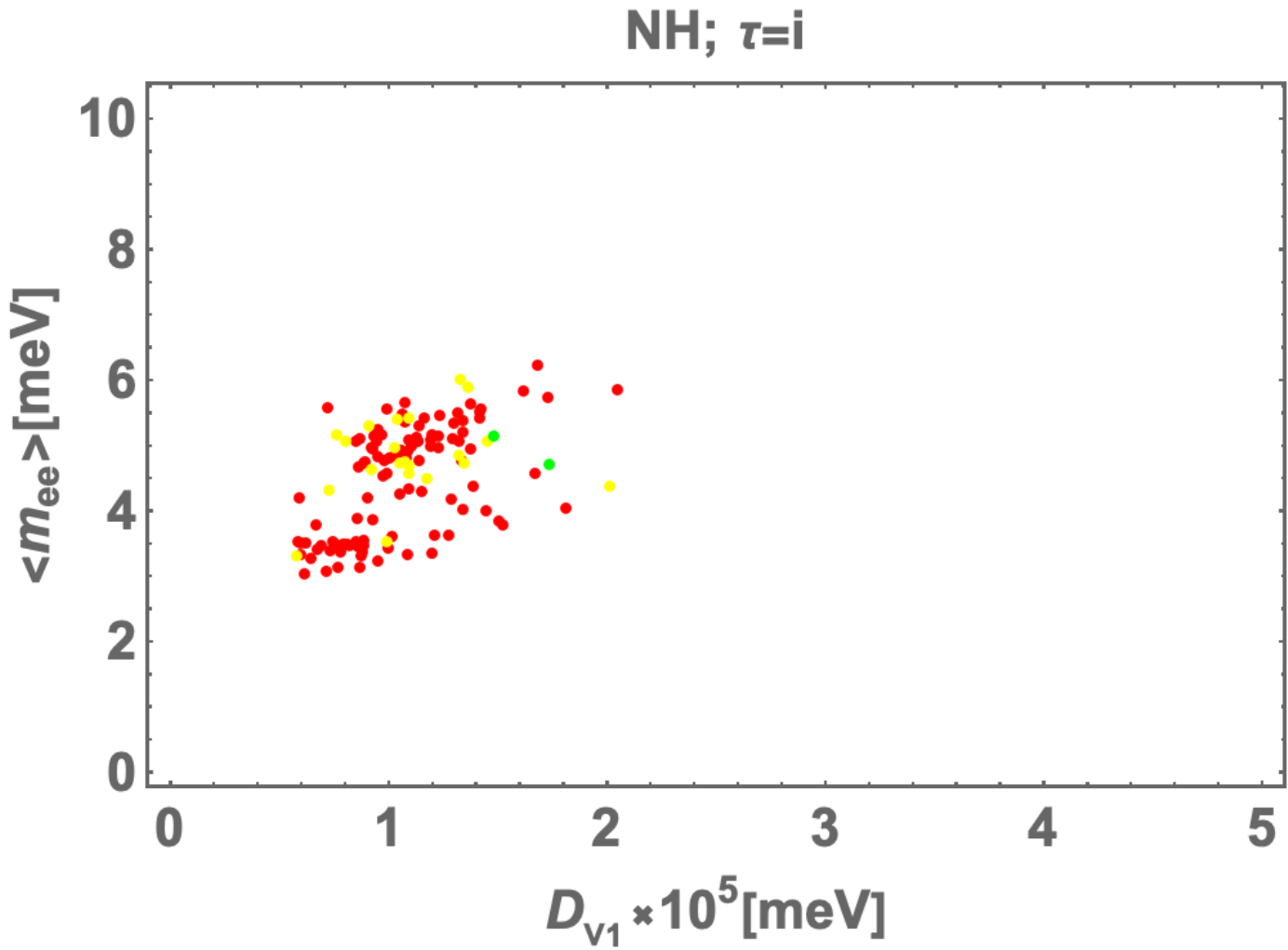} \quad
%%%
\includegraphics[width=50.0mm]{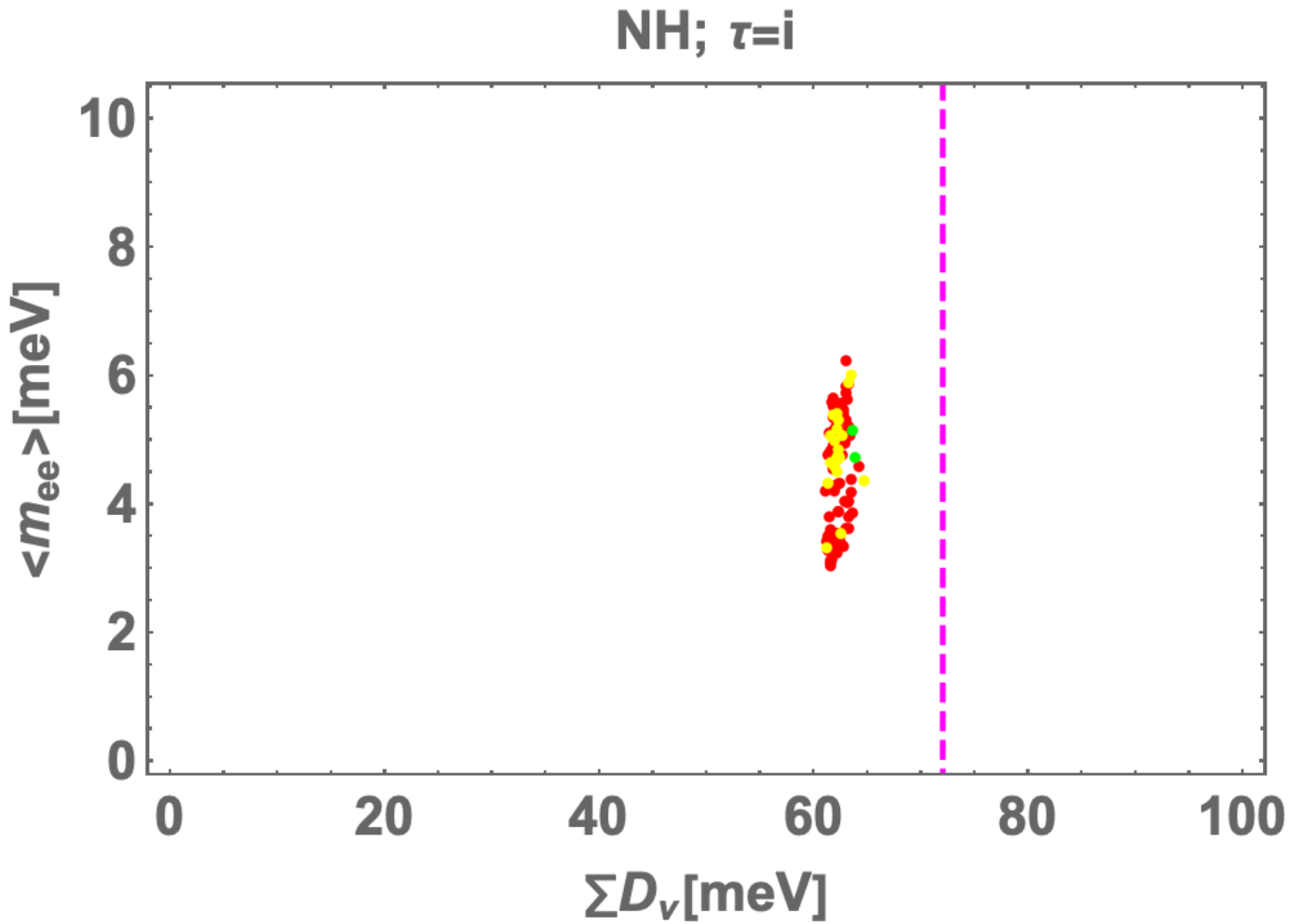} \quad
%%%
\caption{Allowed region of Majorana phases (left), $\langle m_{ee}\rangle$ in terms of the lightest neutrino mass eigenvalue $D_{\nu_1}$ (center), and  $\langle m_{ee}\rangle$ in terms of $\sum D_{\nu}$ (right).
All the color legends are the same as the one of Fig.~\ref{fig:omega_nh2}.}
  \label{fig:i_nh2}
\end{center}\end{figure}
%%%%%%%%%%%%%%%%%%%   
%
In Fig.~\ref{fig:i_nh2}, we show some allowed regions of Majorana phases (left), $\langle m_{ee}\rangle$ in terms of the lightest neutrino mass eigenvalue $D_{\nu_1}$ (center), and  $\langle m_{ee}\rangle$ in terms of $\sum D_{\nu}$ (right).
All the color legends are the same as the one of Fig.~\ref{fig:omega_nh2}.
%The dotted pink vertical line in the right figure shows upper limit of the combination of the DESI and CMB data $72$ meV.
%
In the left figure,
we find  there are two localized islands; $\{ 0^\circ\lesssim\alpha_{21}\lesssim60^\circ,\ 0^\circ\lesssim\alpha_{31}\lesssim80^\circ \}$,
and $\{300^\circ\lesssim\alpha_{21}\lesssim360^\circ,\ 280^\circ\lesssim\alpha_{31}\lesssim350^\circ\}$.
Moreover, Majorana phases are uniquely fixed if either of them is determined due to simple correlation.
In the center figure,
we found $\langle m_{ee}\rangle=[3-6]$ [meV] and  $0.6\times10^{-5}\lesssim D_{\nu_1}\lesssim2\times10^{-5}$ [meV] both of which are so tiny to test by experiments.
%where most of the plots are localized at nearby $D_{\nu_1}=0$ meV.
%
In the right figure,
we found $\sum D_\nu=[60-63]$ [meV] that could be testable in a cosmological experiments/observations such as DESI and CMB data combination near future. 

\subsection{IH}

Here we summarize our findings for IH case.

\subsubsection{$\tau=\omega$}

 %%%%%%%%%%%%%%%%%%%
\begin{figure}[tb]
\begin{center}
\includegraphics[width=50.0mm]{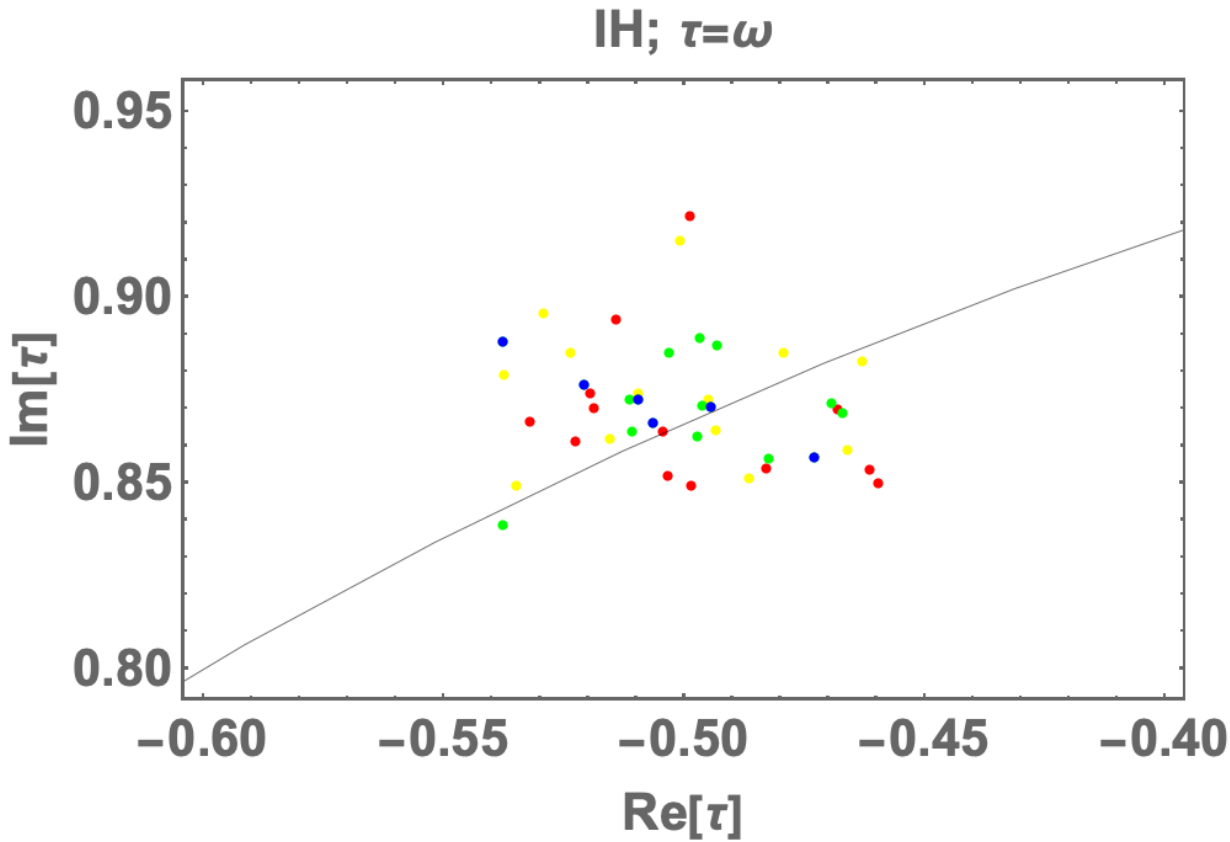}
%%%
\caption{Allowed region of $\tau$ applying $\Delta\chi^2$ analyses in case of IH at nearby $\tau=\omega$.
All the legends are the same as the one of Fig.~\ref{fig:omega_nh1}.    }
  \label{fig:omega_ih1}
\end{center}\end{figure}
%%%%%%%%%%%%%%%%%%%   
%
In Fig.~\ref{fig:omega_ih1}, we show allowed region of Re[$\tau$] and Im[$\tau$] at nearby $\tau=i$ in case of IH.
All the color legends are the same as the one in Fig.~\ref{fig:omega_nh1}.
We find that the allowed region is uniformly located around the fixed point.
%One finds there is deviation only when ${\rm Re}[\omega]\approx0$ or ${\rm Im}[\omega]\approx0$.

 %%%%%%%%%%%%%%%%%%%
\begin{figure}[tb]
\begin{center}
\includegraphics[width=50.0mm]{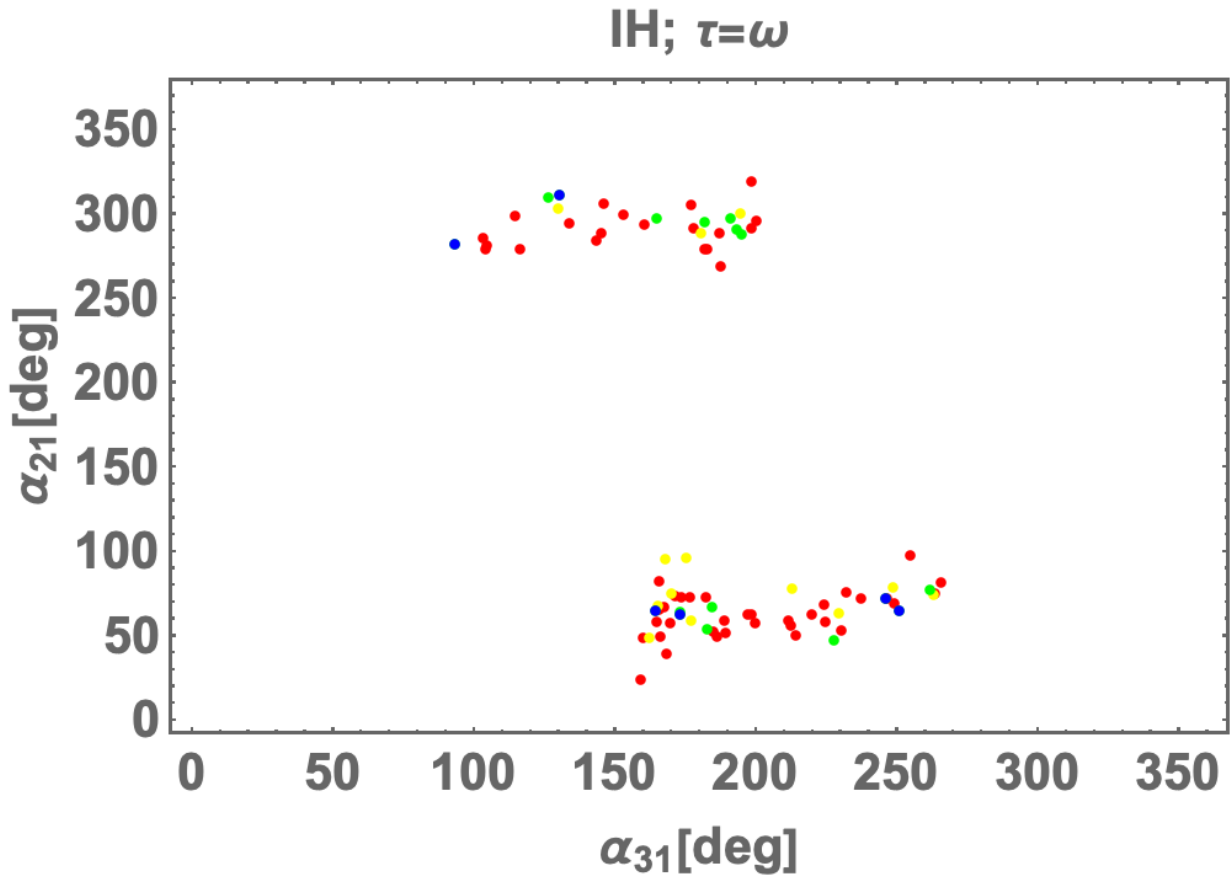} \quad
%%%
\includegraphics[width=50.0mm]{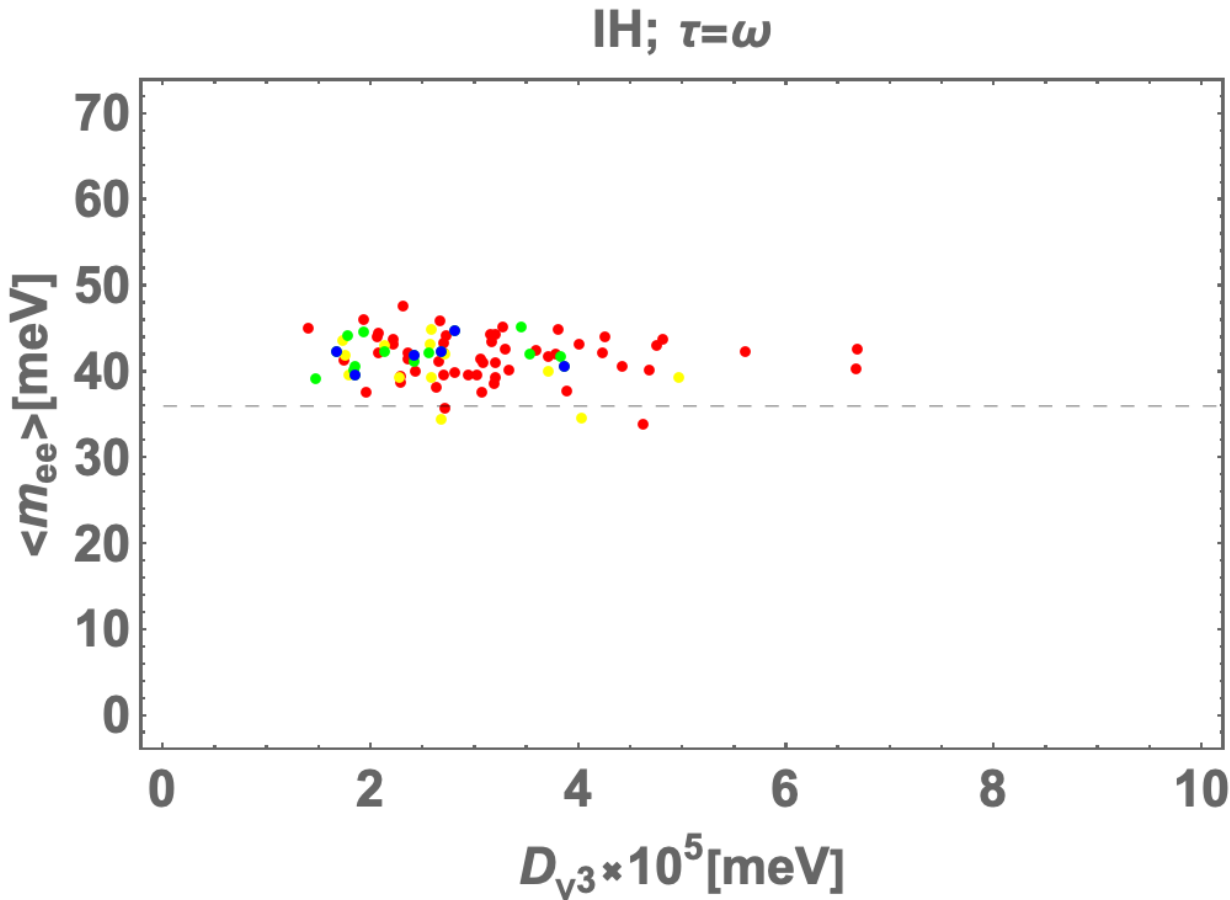} \quad
%%%
\includegraphics[width=50.0mm]{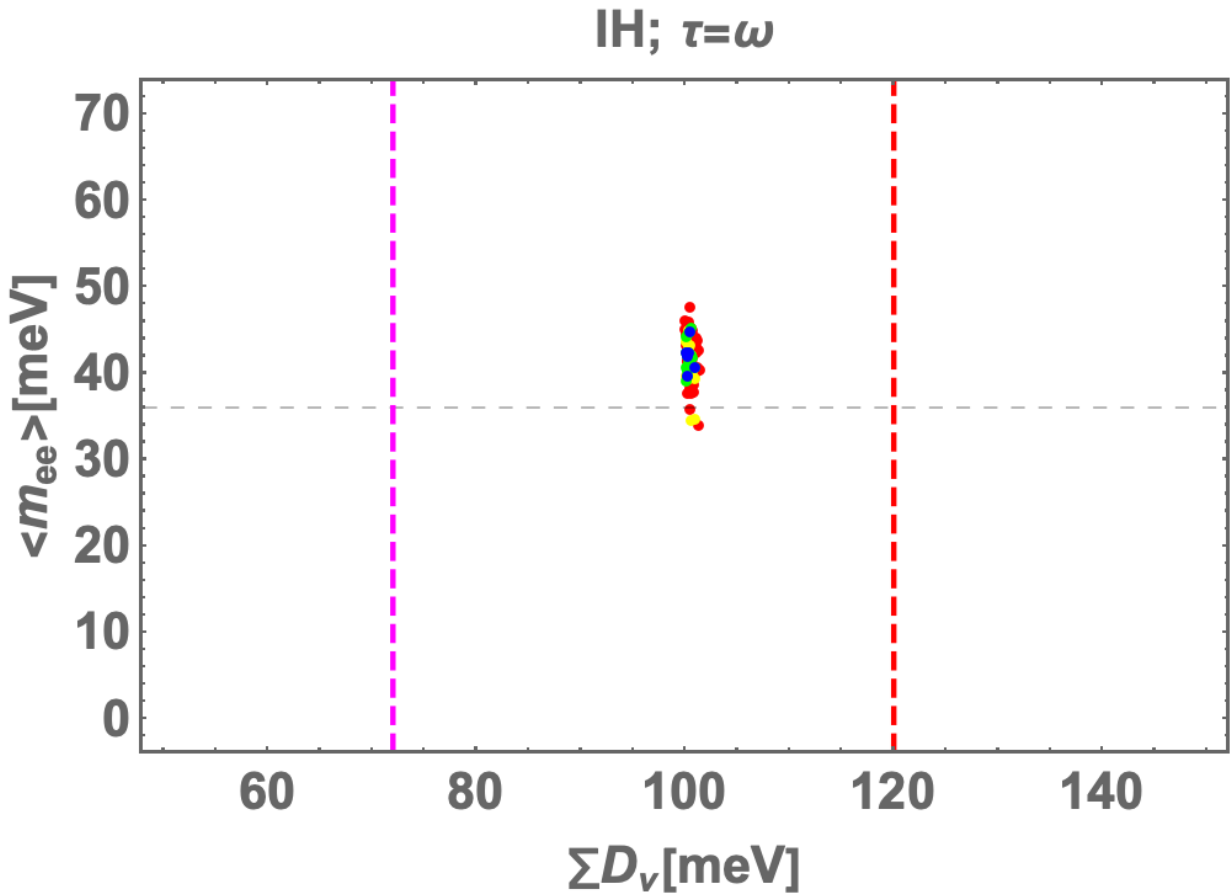} \quad
%%%
\caption{Allowed region of Majorana phases (left), $\langle m_{ee}\rangle$ in terms of the lightest neutrino mass eigenvalue $D_{\nu_1}$ (center), and  $\langle m_{ee}\rangle$ in terms of $\sum D_{\nu}$ (right).
All the legends are the same as the one of Fig.~\ref{fig:omega_nh2}.}
  \label{fig:omega_ih2}
\end{center}\end{figure}
%%%%%%%%%%%%%%%%%%%   
%
In Fig.~\ref{fig:omega_ih2}, we show some allowed regions of Majorana phases (left), $\langle m_{ee}\rangle$ in terms of the lightest neutrino mass eigenvalue $D_{\nu_3}$ (center), and  $\langle m_{ee}\rangle$ in terms of $\sum D_{\nu}$ (right). In the center figure, the horizontal dotted gray line is the lower bound 36 meV.
In the right one, the red vertical line is the cosmological upper bound 120 meV. 
All the other legends are the same as the one of Fig.~\ref{fig:omega_nh2}.
%The dotted pink vertical line in the right figure shows upper limit of the combination of the DESI and CMB data $72$ meV.
%
In the left figure,
{we find}  there are two localized islands; $\{ 20^\circ\lesssim\alpha_{21}\lesssim100^\circ,\ 160^\circ\lesssim\alpha_{31}\lesssim270^\circ \}$,
and $\{270^\circ\lesssim\alpha_{21}\lesssim320^\circ,\ 90^\circ\lesssim\alpha_{31}\lesssim200^\circ \}$.
%Moreover, Majorana phases are uniquely fixed if either of them is determined due to simple correlation.
%
In the center figure,
we found $\langle m_{ee}\rangle=[35-48]$ [meV] and  $1.2\times10^{-5}\lesssim D_{\nu_3}\lesssim6.7\times10^{-5}$ [meV].
Even though the lightest neutrino mass is too tiny, $\langle m_{ee}\rangle$ is localized at nearby the lower bound on 36 meV that would be tested near future. 
%where most of the plots are localized at nearby $D_{\nu_1}=0$ meV.
%
In the right figure,
we found $\sum D_\nu\approx100$ [meV] that could also be testable by cosmological observations.

\subsubsection{$\tau=i$}
%
 %%%%%%%%%%%%%%%%%%%
\begin{figure}[tb]
\begin{center}
\includegraphics[width=50.0mm]{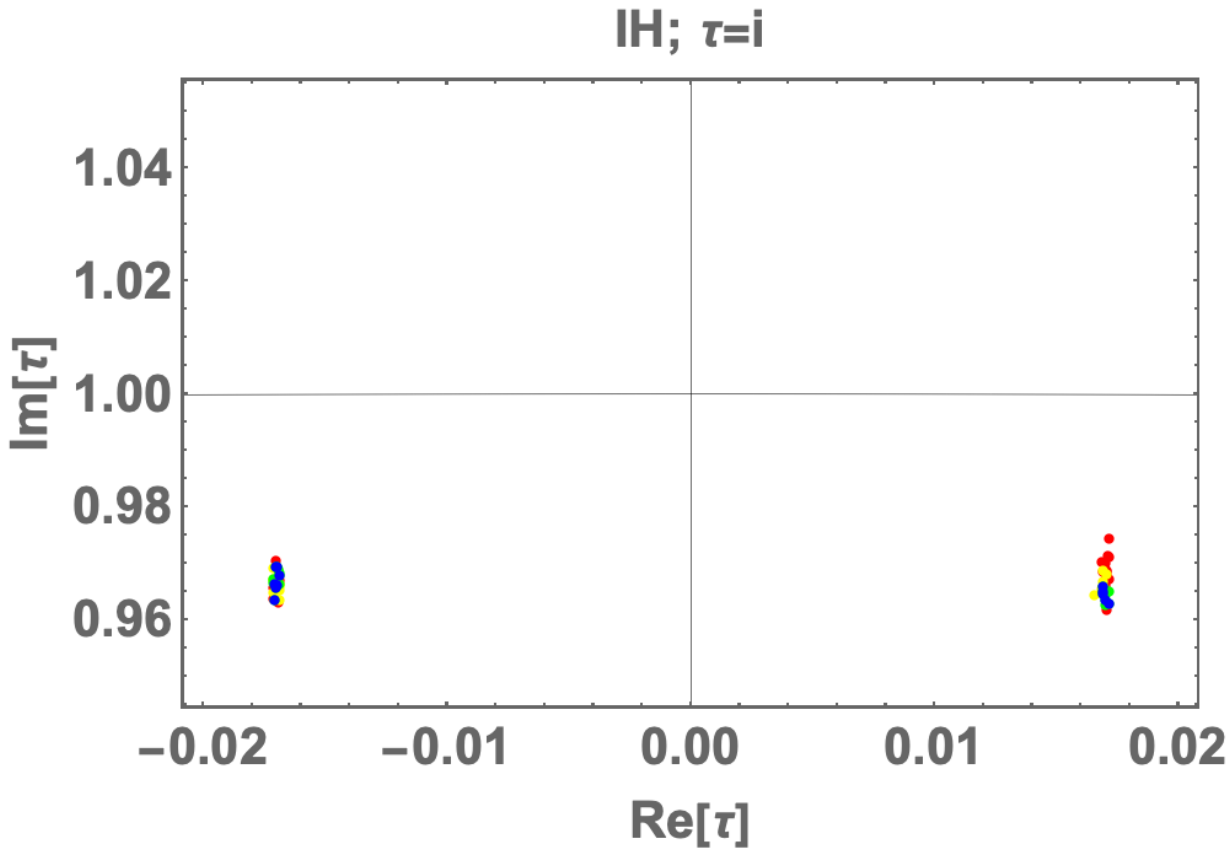}
%%%
\caption{NumAllowed region of $\tau$ applying $\Delta\chi^2$ analyses in case of IH at nearby $\tau=i$.
All the legends are the same as the one of Fig.~\ref{fig:omega_nh1}.     }
  \label{fig:i_ih1}
\end{center}\end{figure}
%%%%%%%%%%%%%%%%%%%   
%
In Fig.~\ref{fig:i_ih1}, we show allowed region of Re[$\tau$] and Im[$\tau$] at nearby $\tau=i$ in case of IH.
All the color legends are the same as the one of Fig.~\ref{fig:omega_nh1}.
We find that the allowed region is localized in two small islands.
%One finds there is deviation only when ${\rm Re}[\omega]\approx0$ or ${\rm Im}[\omega]\approx0$.

 %%%%%%%%%%%%%%%%%%%
\begin{figure}[tb]
\begin{center}
\includegraphics[width=50.0mm]{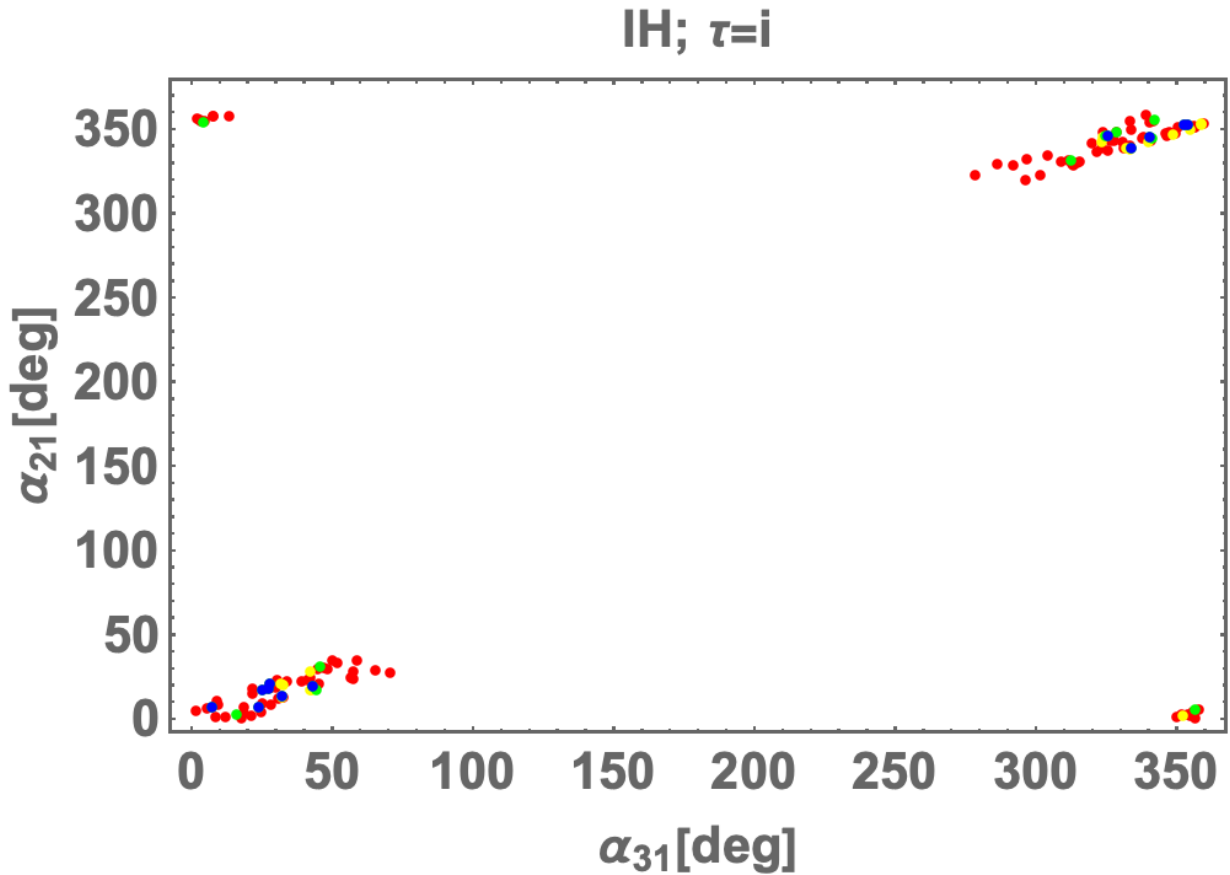} \quad
%%%
\includegraphics[width=50.0mm]{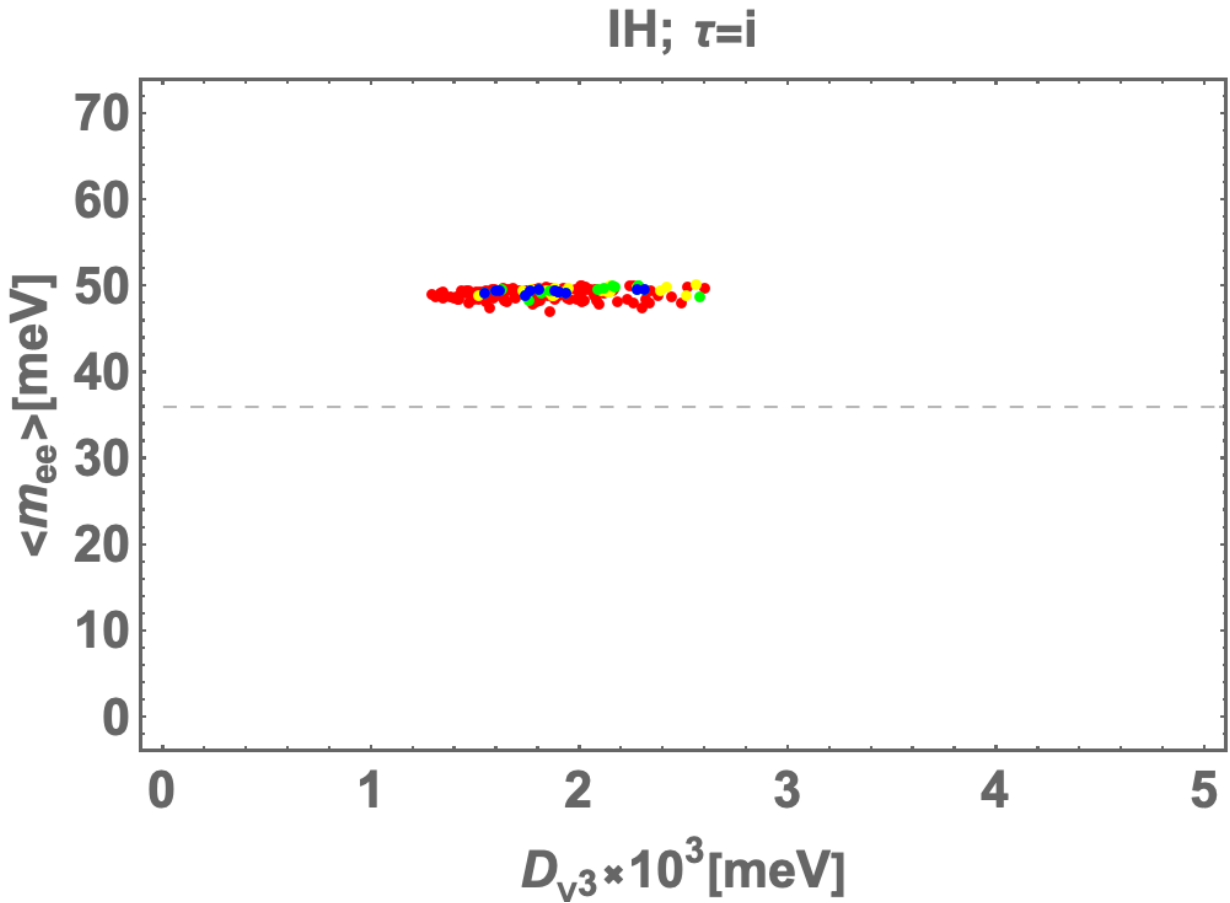} \quad
%%%
\includegraphics[width=50.0mm]{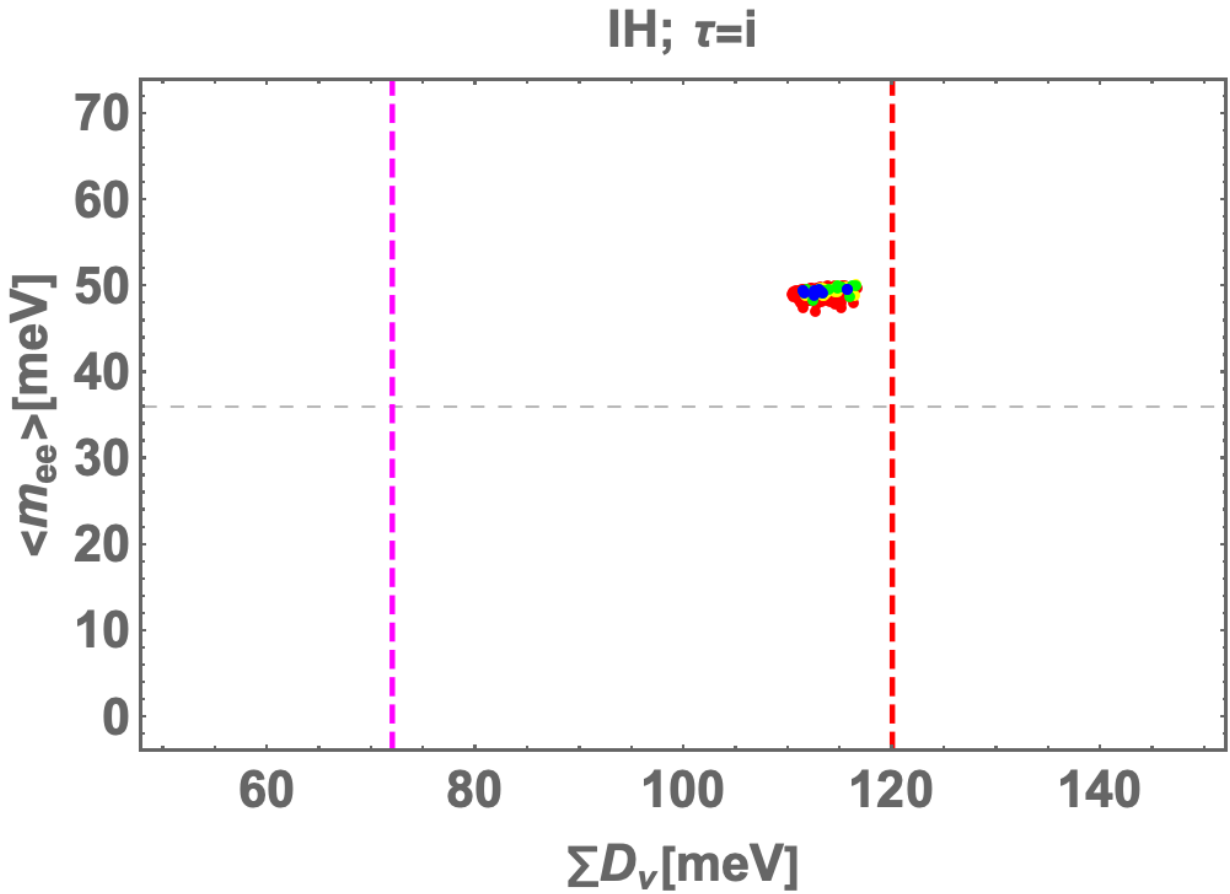} \quad
%%%
\caption{Allowed region of Majorana phases (left), $\langle m_{ee}\rangle$ in terms of the lightest neutrino mass eigenvalue $D_{\nu_1}$ (center), and  $\langle m_{ee}\rangle$ in terms of $\sum D_{\nu}$ (right).
All the legends are the same as the one of Fig.~\ref{fig:omega_nh2}.}
  \label{fig:i_ih2}
\end{center}\end{figure}
%%%%%%%%%%%%%%%%%%%   
%
In Figs.~\ref{fig:i_ih2}, we show some allowed regions of Majorana phases (left), $\langle m_{ee}\rangle$ in terms of the lightest neutrino mass eigenvalue $D_{\nu_3}$ (center), and  $\langle m_{ee}\rangle$ in terms of $\sum D_{\nu}$ (right). In the center figure, the horizontal dotted gray line is the lower bound 36 meV.
In the right one, the red vertical line is the cosmological upper bound 120 meV. 
All the other legends are the same as the one of Fig.~\ref{fig:omega_nh2}.
%The dotted pink vertical line in the right figure shows upper limit of the combination of the DESI and CMB data $72$ meV.
%
In the left figure,
{we find}  there are two larger localized islands; $\{0^\circ\lesssim\alpha_{21}\lesssim40^\circ,\ 0^\circ\lesssim\alpha_{31}\lesssim80^\circ \}$,
and  $\{ 320^\circ\lesssim\alpha_{21}\lesssim360^\circ,\ 270^\circ\lesssim\alpha_{31}\lesssim360^\circ \}$, and two smaller localized islands;
$\{ \alpha_{21}\approx0^\circ,\ 0^\circ\lesssim\alpha_{31}\lesssim20^\circ \}$,
and  $\{ \alpha_{21}\approx0^\circ,\ 350^\circ\lesssim\alpha_{31}\lesssim360^\circ \}$.
%Moreover, Majorana phases are uniquely fixed if either of them is determined due to simple correlation.
%
In the center figure,
we found $\langle m_{ee}\rangle=[48-50]$ [meV] and  $1.2\times10^{-3}\lesssim D_{\nu_3}\lesssim2.6\times10^{-3}$ [meV].
Although the lightest neutrino mass is too tiny, $\langle m_{ee}\rangle$ is close to the lower bound on 36 meV that would be tested near future. 
%where most of the plots are localized at nearby $D_{\nu_1}=0$ meV.
%
In the right figure,
we found $110\lesssim \sum D_\nu\lesssim115$ [meV] that is close to the cosmological upper bound and it could also be testable near future
%

%%%%%%%%%%%%%%%%%%%%
\section{Summary and discussion}
\label{sec:IV}
We have proposed a radiatively induced linear seesaw model perfectly controlling the desired neutral fermion mass matrix at the leading order
through a modular flavor $A_4$ symmetry. {The nature of the symmetry provides a crucial improvement of our previous model in ref.~\cite{Das:2017ski} reducing assumptions to realize linear seesaw mechanism.
%%%
Furthermore, thanks to this modular symmetry, we have obtained some predictions of neutrino observables focusing on specific regions at nearby two fixed points $\tau=\omega$ and $i$.}
In the following, we remark several features in our findings.

Even though the case of $\tau=\omega$ has allowed points almost on $\tau=\omega$,  there is deviation from $\tau=i$ a little for both the cases of NH and IH.
In case of $\tau=\omega$ and $i$ with NH, $\sum D_\nu$ can be testable in a cosmological experiments/observations in near future
although the other mass observables are too tiny. 
In case of $\tau=\omega$ and $i$ with IH, $\langle m_{ee}\rangle$ and $\sum D_\nu$ can be testable in current experiments near future.

%%%%%%%%%%%%%%%%%%%%%%%%%%%%%%%%%%%
\section*{Acknowledgments}
%\vspace{0.3cm}
The work was supported by the Fundamental Research Funds for the Central Universities (T.~N.).
%%%%%%%%%%%%%%%%%%%%%%%%%%%%%%%%%%%

% Ref Style
% Including title
%\bibliographystyle{utphys}
\bibliography{ctma4.bib}
\end{document}